# Molding Free-Space Light with Guided-Wave-Driven Metasurfaces

Xuexue Guo, Yimin Ding, Xi Chen, Yao Duan, and Xingjie Ni*

[1]*Department of Electrical Engineering, Pennsylvania State University, University Park, PA 16802, USA.*

*Correspondence to: xingjie@psu.edu

**Abstract:**

Metasurfaces with unparalleled controllability of light have shown great potential to revolutionize conventional optics. However, they mainly work with free-space light input, which makes it difficult for full on-chip integration. On the other hand, integrated photonics enables densely packed devices but has limited free-space light controllability. Here, we show that judiciously designed guided-wave-driven metasurfaces can mold *guided waves* into *arbitrary free-space modes* to achieve complex free-space functions, such as beam steering and focusing, with ultra-small footprints and potentially no diffraction loss. Based on the same concept together with broken inversion symmetry induced by metasurfaces, we also realized direct orbital angular momentum (OAM) lasing from a micro-ring resonator. Our study works towards complete control of light across integrated photonics and free-space platforms, and paves new exciting ways for creating multifunctional photonic integrated devices with agile access to free space which could enable a plethora of applications in communications, remote sensing, displays, and etc.

With the fast-growing demands for big data, electronic chips and interconnects with insufficient bandwidth can hardly meet the requirements on data transmission speed and energy efficiency of future computing and storage systems. Wiring light on a chip like electronic circuits, integrated photonics provides a promising long-term solution [1,2]. A photonic integrated circuit (PIC) combines many light-controlling components into a single chip, with the ultimate aim of creating miniature optical circuits similar to CMOS (complementary metal oxide semiconductor) chips that have revolutionized the electronics industry. It offers great advantages in terms of speed, bandwidth, reliability, scalability, power consumption, and etc. In order to fully exploit the benefits of PICs in free-space applications, it is crucial to have an interface that can flexibly control light when it converts between guided and free-space modes. However, two conventional coupling techniques – edge couplers [3] and surface gratings [4] – have limited functionalities and lack complete control over light. Although arrays of gratings can achieve more advanced functions, such as off-chip beam steering [5], focusing [6], and holographic image construction [7], they have large footprints and suffer from loss due to the existence of high-order diffractions. Subwavelength gratings [8] have compact footprints but they mainly works in controlling the guided waves rather than the manipulation of light across PICs and free space. Recently, optical nanoantennas are integrated on top of waveguides [9], which provides a new way for interfacing guided and free-space optical modes and adds more functionalities to PICs. Nevertheless, most of them rely on a singular property of the nanoantennas to achieve a specific purpose, such as photo-detection [10] or modulation [11] based on plasmonic field enhancement and directional routing from spin–orbit coupling [12]. Although phase-gradient plasmonic metasurfaces were used for guided mode conversion [13], they only control the phase of guided waves and their phase shifts are limited to π. A unified approach leveraging the collective free-space functions of nanoantennas on PICs has not been demonstrated.

On the other hand, newly emerging metasurface [14,15] – an ultrathin artificial surface which manipulates light by locally imposing abrupt changes to optical properties through engineered sub-wavelength structures also known as meta-atoms – provides unparalleled controllability to the free-space light propagation. However, most of the metasurfaces are driven by free-space light to realize functions, such as beam steering [14,15], generating orbital angular momentum beams [14], light focusing [16], and holograms [17], which makes it difficult for further on-chip integration (e.g. integrating with light sources on the same chip).

Here, we combined synergically two powerful, complimentary technologies: integrated photonics and metasurfaces, and developed a hybrid architecture where metasurfaces are directly driven by guided waves to realize complex free-space functions. We placed subwavelength-sized meta-atoms on top of photonic integrated components (Fig. 1A). In contrast to existing metasurfaces that operate with both input and output light in free space, our integrated metasurface bridges guided waves inside a waveguide with free-space ones. Through it the guided light is tapered into free space and molded into desired light fields. The subwavelength spacing of the meta-atoms eliminates diffraction loss and also allows denser on-chip integration. Meanwhile, multiple metasurfaces can be connected via photonic integrated waveguides to achieve different free-space functions simultaneously. The developed technology will potentially be a huge step towards full control of light across integrated photonics and free-space platforms, and will pave new exciting ways for building multifunctional PIC devices with flexible access to free space as well as guided-wave-driven metasurfaces with full on-chip integration capability. It could enable a plethora of applications in optical communications, optical remote sensing (e.g. light detection and ranging (LiDAR) [18]), free-space optical interconnects (FSOIs) [19], and displays [20]. In addition, a library of those functional hybrid components can be established for reusing and creating consistency across various devices or systems.

We used nano-bar antennas – which evanescently couple with the guided waves inside the waveguides – as the meta-atoms. The fundamental transverse electric mode ($TE_{00}$) in a rectangular waveguide was used to excite resonant modes of meta-atoms as its field distribution has a good spatial overlap with the electric dipolar mode in a nano-bar antenna. However, a single dipolar resonance only provides a phase shift about π at most [21]. In order to have the full 2π phase covered, we introduced a metal-dielectric-metal sandwiched structure (Fig. 2A). The top and bottom of the antenna are two gold cuboids while the middle layer is a dielectric. The thickness of those three layers were carefully chosen so that when the bottom cuboid is excited by the evanescent tail of a guided wave and induced an electric dipole, an antiparallel one can be induced in the top cuboids (Fig. 2A right inset). Those two electric dipoles form a magnetic dipolar resonance. Simultaneously, the magnitude of the electric dipole induced in the top cuboid is smaller than that of the bottom one, thus they do not completely cancel out and result in a net electric dipolar resonance. **The magnetic resonance combined with the electric one creates a directional radiation that extracts the guided wave to free space, and it also provides an abrupt phase shift range ~2π to the extracted wave.** This abrupt phase shift can be tuned by varying the geometrical parameters of the meta-atoms. In addition, by controlling the amount of spatial mode overlap between the antenna mode and the guided mode, we are able to flexibly adjust amplitude of the extracted wave. With the light extraction and control capabilities of meta-atoms, various free-space optical functions can be realized by distributing them along the waveguide. Besides, metasurfaces with different functions can be strategically distributed anywhere on a single waveguide, achieving multiple functions simultaneously, which benefits denser integration. *In* *contrast to the free-space metasurfaces where the spatial phase profile is solely provided by the meta-atoms, the total phase shift of the extracted wave from our guided-wave-driven metasurface is contributed from two parts*: (**i**) the phase accumulation from the propagation of the guided wave and (**ii**) the abrupt phase change induced by each meta-atom (Fig. 1B). It is worth noting that reversing the guided wave propagation direction will change the spatial phase profile and therefore the coupling to free space between the forward and backward guided waves is asymmetric (Fig. 2D).

We simulated the radiation phase shift with respect to the guided wave right underneath the meta-atom using full-wave finite element method (FEM) (supplementary materials). We chose meta-atom designs that have a uniform amplitude of the extracted wave while having phase shifts that cover entire $2\pi$ phase range, as indicated by the white stars in Fig. 2B. The simulated electrical field distribution for the three selected meta-atoms on the waveguide shows that the extracted waves indeed have distinct abrupt phase shifts that covers a $2\pi$ range (Fig. 2C).

In order to show the capability of the guided-wave-driven metasurfaces, we numerically and experimentally demonstrated off-chip beam steering and light focusing directly from a photonic integrated waveguide. Besides, by spatially arranging meta-atoms on PICs with optical gain materials, we created micrometer-sized photonic integrated micro-ring lasers that directly emit vector optical vortices carrying well-defined, quantized OAM.

**Off-Chip Beam Steering with an Integrated Guided-Wave-Driven Metasurface**

The extracted light from the guided-wave-driven metasurface carries phase consisting of accumulated propagation phase $\beta x$ (where $\beta$ is the propagation constant of the guided mode and we assumed that the waveguide lays straight along the $x$ direction) and the induced abrupt phase shift $\Delta\phi(x)$ by the meta-atom at coordinate $x$ (Fig. 1B). As a result, the phase distribution of the extracted wave along the $x$ direction can be expressed as

$$\phi(x) = \beta x + \Delta\phi(x), \quad (1)$$

Therefore, linear momentum of extracted light along the $x$ direction is $k_x = \beta + \partial\Delta\phi(x)/\partial x$. If $\partial\Delta\phi(x)/\partial x$ is a constant, the extracted beam has a well-defined angle $\theta$ given by $\theta = \sin^{-1}(k_x/k_0)$, where $k_0$ is the free-space wave number. In contrast to grating couplers on PICs, our metasurface approach introduces abrupt and large phase shifts with a subwavelength spacing, which eliminates high-order diffractions and offers a much large beam steering angle range.

We used Au/$SiO_2$/Au sandwich-structured nanoantennas as the meta-atoms for beam steering. The meta-atoms are periodically distributed on a silicon waveguide to provide a phase gradient $\partial\Delta\phi(x)/\partial x = -2\pi/\Lambda$, where $\Lambda$ is the length of a supercell which consists of three meta-atoms with abrupt phase shifts $-2\pi/3$, 0, and $2\pi/3$. Therefore, the output angle of the extracted beam is $\theta = \sin^{-1}\frac{1}{k_0}\left(\beta - \frac{2\pi}{\Lambda}\right)$ (Fig. 2D). The asymmetric coupling effect can be seen by reversing the propagation direction of the guided wave. The resulting momentum along the $x$ direction becomes $k_x = -\beta - 2\pi/\Lambda$ which is too large to be supported in free space (Fig. 2E). In this case, the extracted wave bounds to the metasurface and eventually dies out due to ohmic loss from the materials. It is worth noting that reciprocity of the system is preserved as the transmitted power in the waveguide for the forward- and backward-propagating waves is equal.

We fabricated the beam steering samples using two electron beam lithography steps with precise alignment to define the silicon waveguide and the meta-atoms (Fig. 3A) Different lengths of supercells were chosen to demonstrate flexible control of the beam steering angles. The propagation constant $\beta$ was numerically calculated for the fundamental TE modes at different wavelengths. *Fourier-space* imaging system was employed to measure the scattering angles. We experimentally measured the output angles with different wavelengths and supercell periods, respectively, and the results agree well with those from our theoretical calculations (Fig. 3B and C). The slight discrepancy originates from the fabrication error. The line-shaped intensity profile in Fourier space reveals the in-plane wavevector of the extracted light, where $k_x$ is determined by the metasurface and $k_y$ spans the whole Fourier plane because no phase modulation is applied in y direction. The divergence of the steering angle, which is depicted by the width of the line, is inversely related to the length of the metasurface region. The bright ends of the lines (near the cut-off lines in Fourier space images limited by the numerical aperture (NA = 0.95) of the objective) are originated from the internal reflection in the objective. The Fourier space images were also validated by theoretical calculations (Fig. S2).

**Off-Chip Light Focusing with an Integrated Guided-Wave-Driven Metalens**

By spatially arranging the meta-atoms along a waveguide to fulfill a lens phase function $\phi(x) = -k_0\sqrt{x^2 + f^2}$, we can focus the wave in free space with a designated focal length *f*. Therefore, combining with Eq. (1) the abrupt phase provide the meta-atoms should be

$$\Delta\phi(x) = -k_0\sqrt{x^2 + f^2} - \beta x. \qquad (2)$$

As a proof of concept, we simulated such a metalens on a silicon waveguide with a focal length *f* = 5 μm (we chose a short focal length in order to reduce the demand for computational resources) at 1550 nm (Fig. 4A). Evidently, light is extracted and focused into free space by the metalens. We designed and fabricated a larger guided-wave-driven metalens with a focal length of 225 μm.

We obtained the intensity distribution at different heights above the waveguide and reconstructed the intensity profile in the $xz$ plane (Fig. 4B). It shows a similar focusing effect as that in our simulation.

## Photonic Integrated OAM Lasers

Leveraging the asymmetric coupling induced by the guided-wave-driven metasurface, we are able to create a photonic integrated micro-ring OAM laser (Fig. 5A). Light beams with an azimuthal phase profile of the form $\exp(il\varphi)$ carry an OAM of $l\hbar$ [22], where $l$ is an integer known as the topological charge, and $\varphi$ is the azimuthal angle with respect to the propagation direction. Light can have infinite number of orthogonal OAM states essentially. This unique property makes it an excellent candidate for encoding information in both classical [23,24] and quantum [25,26] optical communications as well as many other applications [27]. A conventional system for generating OAM light usually have two separate parts – a light source and an optical component for spatial phase modulation, e.g. spatial light modulator [28], phase plates [29,30], and metasurfaces [31], which makes it bulky, poor in scalability, and difficult for on-chip integration. An compact, integratable, and scalable source directly emits OAM light [32,33] is highly desirable.

The micro-ring resonator intrinsically supports two degenerate whispering gallery modes (WGMs) – a clockwise (CW) and a counter-clockwise (CCW) mode. These modes by themselves carry high-order OAM. But due to the inversion symmetry of the micro-ring, the OAMs of the CW and CCW modes have opposite signs and the net OAM is zero [33]. In order to obtain controllable OAM emission, our metasurface accomplished three things: (**1**) Extract light from the micro-ring without destroying the guided modes; (**2**) Break the degeneracy of the two WGM modes to get non-zero net OAM emission; (**3**) Control the topological charge of the OAM.

Due to the **asymmetric coupling** effect of the guided-wave-driven metasurface, only one of the two counter-propagating WGMs can couple to the free-space emission, and therefore **we are able to break the degeneracy of the WGMs and achieve a controllable OAM emission.** As the degenerate WGMs interact with the metasurface on a micro-ring that introduces a unidirectional phase gradient $\mathrm{d}\phi/\mathrm{d}\varphi$ ($\varphi$ is the azimuth angle), the radiated light of CW and CCW mode will gain additional but opposite momenta. One radiation mode will gain too large $k$ to propagate in free space, while the other one can be successfully launched into free space with a well-defined OAM order.

Let us suppose we want the CCW mode to be extracted and form OAM emission in free space. The propagation constant of the $M^{\text{th}}$-order CCW WGM is given by $\beta_{\text{CCW}} = 2\pi n_{\text{m}}/\lambda = M/R$, where $n_{\text{m}}$ is the modal index and $R$ is the micro-ring radius. The guided-wave-driven metasurface is placed on the micro-ring so that it induces a phase gradient that is equivalent to a wave number $k_{\text{ms}} = -2\pi/\Lambda$, where the phase shifts provided by the meta-atoms decreases linearly along the CCW direction and $\Lambda$ is the size of the metasurface supercell. The azimuthal phase dependence of OAM emission can be expressed as $\phi_{OAM}(\varphi) = l\varphi$. Due to momentum conservation, the following condition should be satisfied

$$l\varphi = \phi_{OAM} = \phi_{CCW} + \phi_{ms} = \beta_{\text{CCW}} R\varphi - \frac{2\pi}{\Lambda} R\varphi. \tag{3}$$

Assuming the total number of metasurface supercells on the micro-ring is $N = \frac{2\pi R}{\Lambda}$, we can obtain from Eq. (3) a well-defined topological charge

$$l = M - N, \tag{4}$$

which can be easily engineered either by designing the order of the WGM mode or by placing different numbers of supercells on the micro-ring.

We designed our OAM laser based on an InGaAsP/InP multi-quantum-well (MQW) micro-ring resonator. Four Au/Si/Au sandwich-structured meta-atoms covering $2\pi$ abrupt phase shift range (Fig. S3A) were used to construct one metasurface supercell and patterned periodically on top of the micro-ring (Fig. 5A). We showed using full-wave FEM eigen-mode simulations that the emitted light is radially polarized and exhibits the characteristics of OAM emission. With $M = 59$ and $N = 58$, the electric field $E_r$ forms a spiral pattern, and its phase changes by $2\pi$ upon one full circle around the center of the vortex, indicating $l = 1$ (Fig. 5B, top row). We also showed the simulation results of $l = 2$ with $M = 59$ and $N = 57$ (Fig. 5B, bottom row). The phase profile depicts a $4\pi$ winding around the center of the vortex.

We fabricated the micro-ring OAM laser (Fig. 5B) and characterized its lasing properties. The micro-ring was pumped by 900-nm femtosecond pulses (~140 fs) from a Ti:Sapphire laser, and the radiation from the micro-ring was collected and analyzed by a spectrometer (Fig. S4). The spectra gradually transitioned from spontaneous emission (SE) to amplified spontaneous emission (ASE) and finally to lasing as the pump intensity increased (Fig. 5D). The OAM characteristics were characterized by analyzing both the spatial intensity profile of the emission using a near-infrared camera and its self-interference pattern using Michelson interferometry (Fig. S4). We observed the intensity of lasing emission spatially distributed in a doughnut shape with a dark core in the center (Fig. 5E), which is due to the phase singularity at the beam axis where the phase becomes discontinuous. The presence of the OAM was also validated by the self-interference patterns (Fig. 5 F and G). We split equally the beam emitted from the micro-ring into the two arms of a homebuilt Michelson interferometer. Because in an OAM beam, the phase varies drastically (helically) close to the central singularity, whereas it is relatively uniform (quasi-planar) at the outer rim, we intentionally created a horizontal offset between the two split OAM beams at the observation plane, so that the dark center of one beam overlapped with the bright outer rim of the other, and vice versa. The interference between helical and quasi-planar phase distributions revealed two inverted forks in the resulting fringes (Fig. 5F). In each folk, a single fringe split into two, which evidently confirmed that the emission from the laser carries OAM with topological charge $l$ = 1. Similarly, OAM laser emission with topological charge of 2 was also observed experimentally in another design (Fig. 5G), which matches perfectly with our theory.

**Discussions**

The guided-wave-driven metasurface, consisting of subwavelength-spaced meta-atoms, provides a highly versatile and compact platform for bridging the gap between guided waves in PICs and free-space waves. The developed technology not only empowers the photonic integrated devices with agile free-space light controllability in the subwavelength scale, but also enables metasurfaces to be directly driven by guided waves which makes possible a denser and higher level of on-chip integration. Recently, waveguide-fed metasurfaces for beam forming and wavefront shaping have been demonstrated in microwave frequencies[34]. Our work well distinguishes itself with a complete control of phase (over $2\pi$) and amplitude in the optical regime. Also, although a waveguide consisting of a perfect electric conductor plate and a bianisotropic Huygens metasurface can support up to $2\pi$ phase tuning for the leaky wave [35], it so far only works in the microwave regime, and its bianisotropic metasurface requires multiple antenna layers, which is challenging to be realized in the optical frequencies. Our work represents a first attempt at a unified designed, guided-wave-driven metasurface in the optical regime.

We have experimentally demonstrated off-chip beam steering and focusing on silicon waveguides using the guided-wave-driven metasurfaces, which could enable a wide spectrum of applications ranging from optical communications to LiDAR, as well as miniaturized display technology for virtual reality (VR) and augmented reality (AR) devices. Taking advantage of the intrinsic asymmetric coupling originated from unidirectional phase distribution provided by the metasurface, we also demonstrated an on-chip micro-ring OAM laser which directly emits beam that carries OAM with a designable order. This technique holds great promise for achieving compact on-chip OAM light sources (or detectors) for large-scale photonic integration. Especially, it can be used for free-space optical communications with an additional degree of freedom provided by the OAM states. Based on the demonstrated design principles, more complex functionalities can be achieved, such as guided-wave-driven holograms, photonic integrated spectrometers and so on. In addition, due to reciprocity, free-space modes can be selectively coupled into the metasurface-dressed waveguides. The metasurface region can be engineered to couple light with a tilted or even distorted wave front into a waveguide, which is especially useful for optical sensing and detections. Moreover, dynamic control of the coupling between guided modes and free-space ones can be realized by incorporating tunable elements [36,37], which further empowers the PICs with the capability of tuning the optical functionalities dynamically.

## Methods

### Sample fabrication

The samples were fabricated on a commercially available silicon-on-insulator wafer with 220-nm-thick (for beam steering experiments) and 500-nm-thick (for light focusing experiments) Si device layer and 3-μm buried silicon dioxide. Alignment marker was defined by electron beam lithography followed by evaporation of 50 nm Au with a 5-nm-thick Ti adhesion layer and lift-off process. Then negative resist Fox-16 (Dow Corning Corp.) was used to define the waveguide pattern and then developed in CD-26 developer (MicroChem) for 25 minutes. Chlorine-based inductively coupled reactive ion etching (ICP-RIE) was used to etch crystalline Si with FOX-16 resist as mask. Then, the sample was immersed in buffered oxide etchant for 20 seconds followed by water rinse to remove the remaining mask. A second-step electron beam lithography was conducted on ZEP 520A (Zeon) resist to define the metasurface on top of the waveguide with precise alignment. The exposed sample was developed in N-amyl-acetate for 3 minutes followed by MIBK:IPA immersion for 1 minute. Au/SiO2/Au films were subsequently deposited using an electron beam evaporation system. The pattern was then lifted off in 1165 remover (MicroChem) at 85 °C in water bath for 2 hours. The sample was finally diced along the input port of waveguide for measurement.

The OAM micro-ring laser was fabricated on InGaAsP (500 nm, multi-quantum-well layer) / InP substrate. First the micro-ring resonator was defined by electron beam lithography with FOX-16 negative resist. The resist acted as an etch mask in the $BCl_3$ based ICP-RIE process. Then the sample was immersed in buffered oxide etchant to remove the mask. A second-step electron beam lithography using ZEP 520A resist was performed with precise alignment to define the metasurface on top of the micro-ring resonator. A sequential electron beam evaporation was done to deposit Au/Si/Au films, followed by a standard lift-off process in 1165 remover at 85 °C in water bath for 2 hours.

**Simulation methods**

Numerical simulations were carried out using a commercially available finite element method (FEM) solver package – COMSOL Multiphysics. Third-order finite elements and at least 10 mesh steps per wavelength were used to ensure the accuracy of the calculated results. We simulated individual Au/$SiO_2$/Au meta-atoms first. We used an eigen-mode solver to find the $TE_{00}$ mode of the silicon waveguide as well as its modal index at 1550 nm wavelength. Then this modal index was used in the model to further calculate the phase and amplitude of the extracted light by monitoring the field at a few wavelengths over the waveguide. We swept the geometrical parameters of the meta-atoms to get the phase and amplitude maps/contours (Fig. 2B). The trapezoidal shape of meta-atoms resulted from our nanofabrication was also taken into account in our model to get accurate design parameters.

In order to simulate the beam steering, a full device model that consists of an array of meta-atoms placed on top of Si waveguide was established. The meta-atoms were distributed along the waveguide so that they formed a linear phase gradient (Fig. 2D). In a similar fashion, an array of meta-atoms fulfilling the spatial phase distribution of a lens were placed on top of Si waveguide to simulate the light focusing effect (Fig. 4A).

We used a similar method to calculate the phase and amplitude of the extracted light from Au/Si/Au meta-atoms on top of InGaAsP/InP waveguide (Fig. S3A). A device-level model of metasurface incorporated micro-ring resonator was constructed to simulate the OAM radiation. Four Au/Si/Au meta-atoms selected from calculated phase and amplitude maps/contours (Fig. S3A) were used to construct a supercell. 58 supercells ($N = 58$) were placed on top of the resonator. Using the WGM with $M = 59$, we achieved OAM radiation with topological charge of +1 according to Eq. (4). Fig. S3B illustrates the electric field distribution of $TE_{00}$ mode which shows the typical standing wave pattern formed by the two counter-propagating (CW and CCW) WGMs. The extracted light in free space carries OAM as shown in Fig. 5B.

**Experimental setups**

We characterized our guided-wave-driven metasurfaces with beam steering and focusing functions using the optical setup shown in Fig. S1. A free-space laser beam output from a Ti:Sapphire laser pumped optical parametric oscillator (OPO) was coupled into a commercially available tapered lensed single-mode fiber. The focused laser beam from the tapered fiber was coupled into the input port of our fabricated ridge waveguide sample in end-fire manner by using a three dimensional (3D) translational stage. The coupled-in light propagated through a triangle taper linking the input port and the single-mode waveguide, during which the high-order modes vanished and only fundamental transverse electrical mode survived. The light scattered into free space by metasurfaces on top of the single-mode waveguide was collected by an objective (N.A. = 0.95) and then transmitted through a tube lens. Part of the light was reflected by a beamsplitter for real-space imaging. And the light transmitted through the beamsplitter was focused by a Bertrand lens to form *Fourier-space* image. In off-chip beam-steering measurement, the laser wavelength was tuned using the OPO to acquire wavelength-dependent beam steering angles. In addition, Fourier-space images were taken by coupling 1550nm laser beam into samples with different supercell periods. After that, the scattering angles were extracted from the Fourier-space images calibrated by a ruled reflective grating (grooves density of 600/mm). In off-chip focusing measurement, the samples were mounted on a 3D translational stage with a high-resolution piezo-controlled actuator in *z*-direction. By moving the *z* stage, the real-space images were taken at different distance from the waveguide plane and a 3D intensity distribution was reconstructed.

To observe the lasing spectra and to confirm the OAM properties of the laser radiation, we used the set-up shown in Fig. S4. In our experiment, a femtosecond pulsed laser (~140 fs, repetition rate 80 MHz) at 900 nm wavelength was reflected by a dichroic mirror and then focused by a Newport 20X objective (NA = 0.40) onto the micro-ring resonator. The pump power was controlled by a circular variable neutral density filter and monitored by a power meter. The lasing emission was collected by the same objective and then transmitted through the dichroic mirror and detected by a spectrometer (Horiba), a far-field imaging system and a Michelson interferometry setup. With a flip mirror to switch the paths, the laser emission was either sent into the spectrometer/imaging system or the interferometry setup. In the interferometry setup, the laser emission was split into two beams by a pellicle beam splitter, and then recombined with an off-center beam overlap to form an interference pattern recorded by an infrared camera. A delay line was used in order to balance the optical path lengths of the two arms.

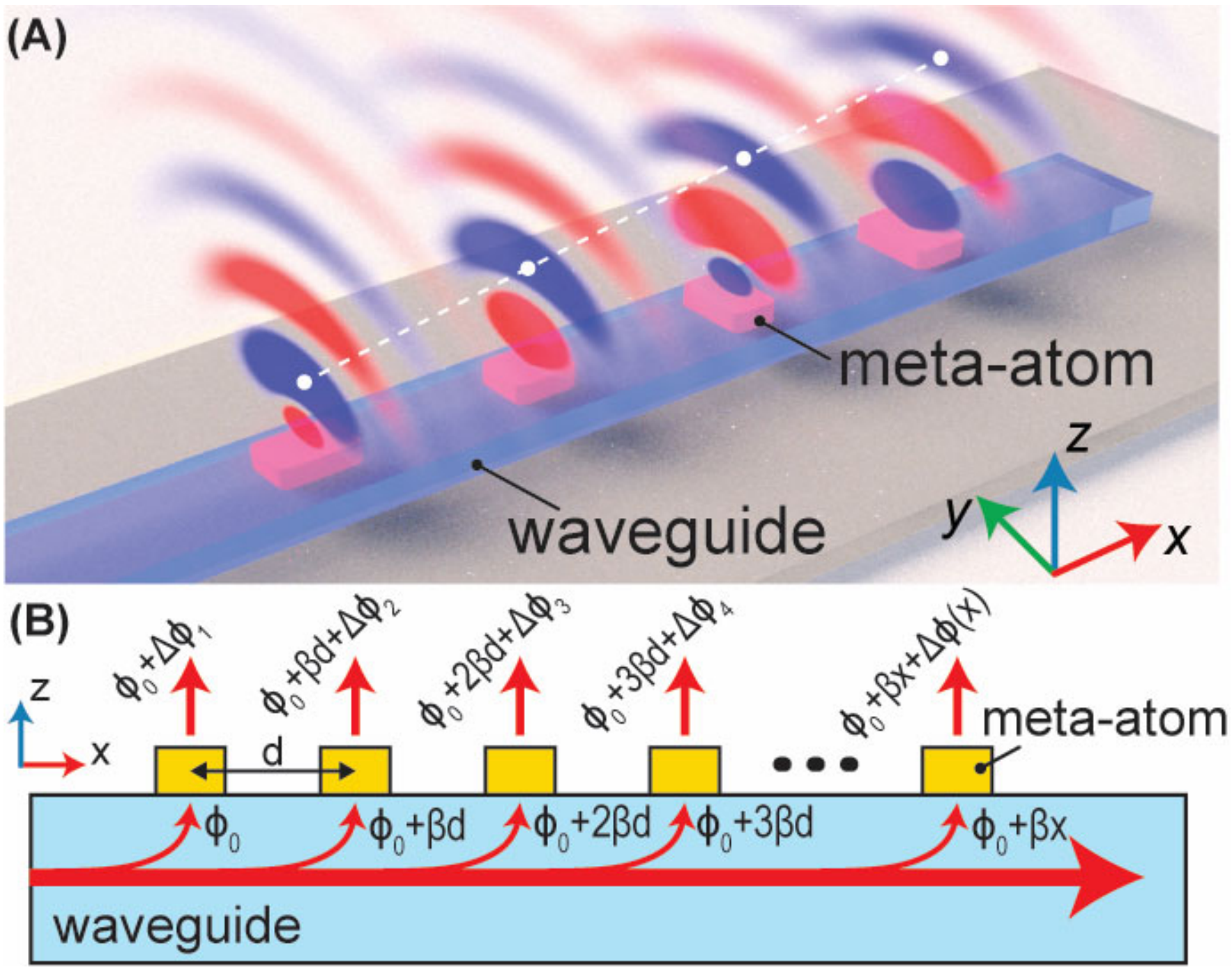

**Fig. 1. The working principle of guided-wave-driven metasurfaces.** (**A**) A schematic of a guided-wave-driven metasurface. The phase of the extracted light from a guided wave by each meta-atom can be tuned individually. An array of meta-atoms on the waveguide work collaboratively to form certain wave fronts and fulfill different functions, such as beam steering and focusing. (**B**) An illustration of the wave front formation of the extracted wave. The total phase shift of the extracted wave at coordinate $x$ is contributed from two parts: the phase accumulation $\beta x$ from the guided wave propagation and the abrupt phase change $\Delta\phi(x)$ induced by the meta-atom. As a result, the phase of the extracted wave can be expressed as $\phi_0+\beta x+\Delta\phi(x)$, where $\phi_0$ is the initial phase of the incidence.

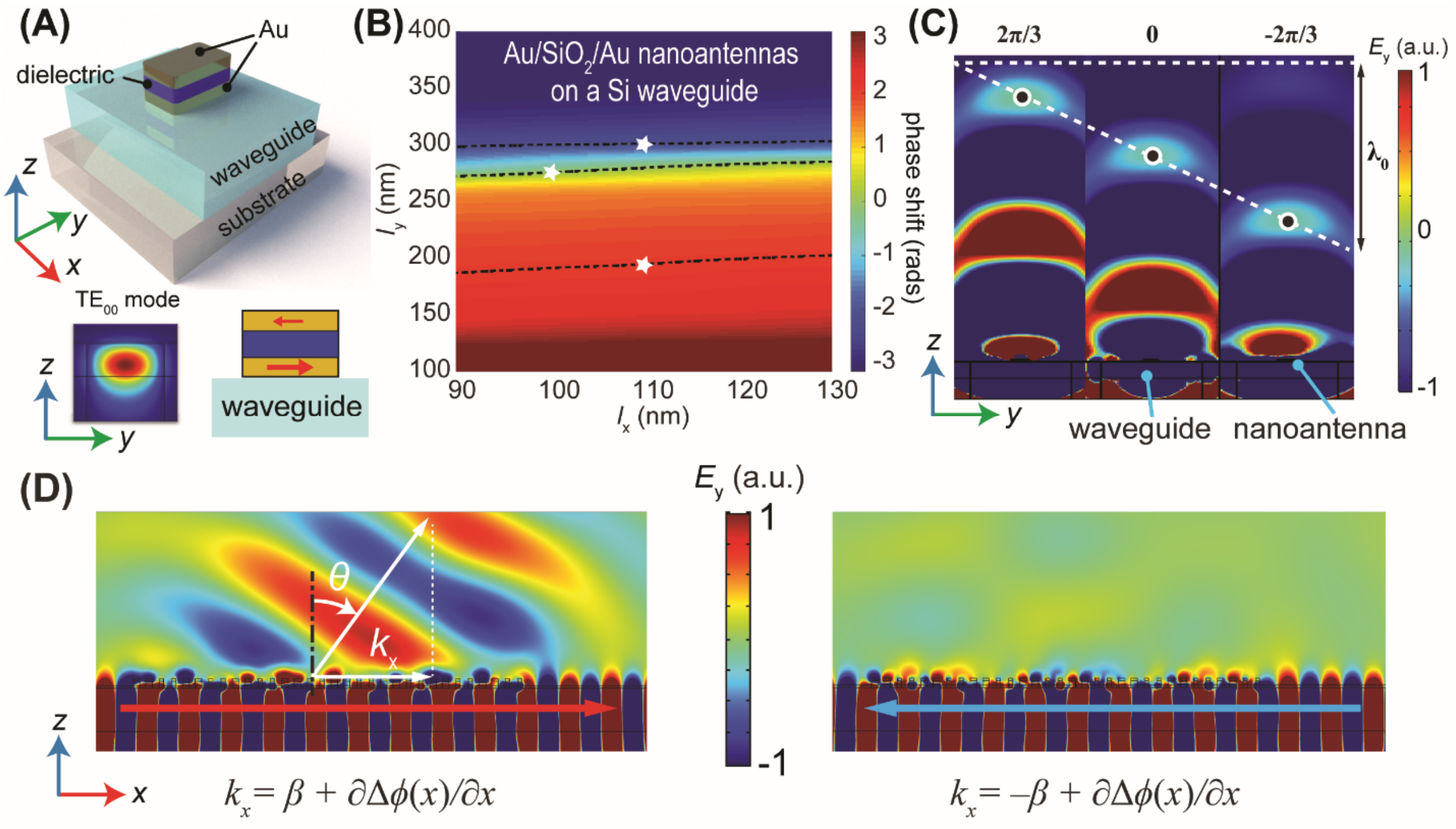

**Fig. 2. Design of meta-atoms for controlling the phase (and amplitude) of free-space waves extracted from guided ones.** (**A**) A schematic of a metal/dielectric/metal sandwich-structured meta-atom on top of a photonic integrated waveguide. The bottom left inset shows the simulated electric field distribution of the $TE_{00}$ guided mode propagating inside the waveguide. The bottom right inset is a conceptual view of the induced displacement current inside the meta-atom, which forms an effective magnetic dipole and a net electric dipole. (**B**) A pseudo-color map of the simulated abrupt phase shifts in a parameter space spanned by the meta-atom width ($l_x$) and length ($l_y$). A thickness of 30 nm was used for each layer. The meta-atom was placed on top of a silicon ridge waveguide (height 220 nm). The three white stars indicate the meta-atom designs covering 2π phase range with an even interval. We also ensured that the extracted waves from the chosen meta-atoms have roughly the same amplitude. (**C**) Simulated electric field distribution ($E_y$) of the extracted waves from the three selected meta-atoms, showing abrupt phase shifts of 2π/3, 0, and −2π/3, respectively. (**D**) Electric field distribution of the extracted light from a phase-gradient

metasurface driven by forward- (left panel) and backward- (right panel) propagating guided waves. The metasurface consists of an array of meta-atoms that form a phase gradient $\partial\Delta\phi(x)/\partial x$ (which is along the $-x$ direction in this example). The extracted light from a forward-propagating guided wave carries a transverse wavevector $k_x = \beta+\partial\Delta\phi(x)/\partial x$, where $\beta$ is the propagation constant of the guided wave. It is launched into free space with a well-defined angle $\theta = \sin^{-1}(k_x/k_0)$. In contrast, light extracted from the backward-propagating wave gains a transverse wavevector so large that it exceeds the maximum supportable wavenumber in free space, and therefore it bounds to the metasurface and eventually dies out due to ohmic loss from the materials.

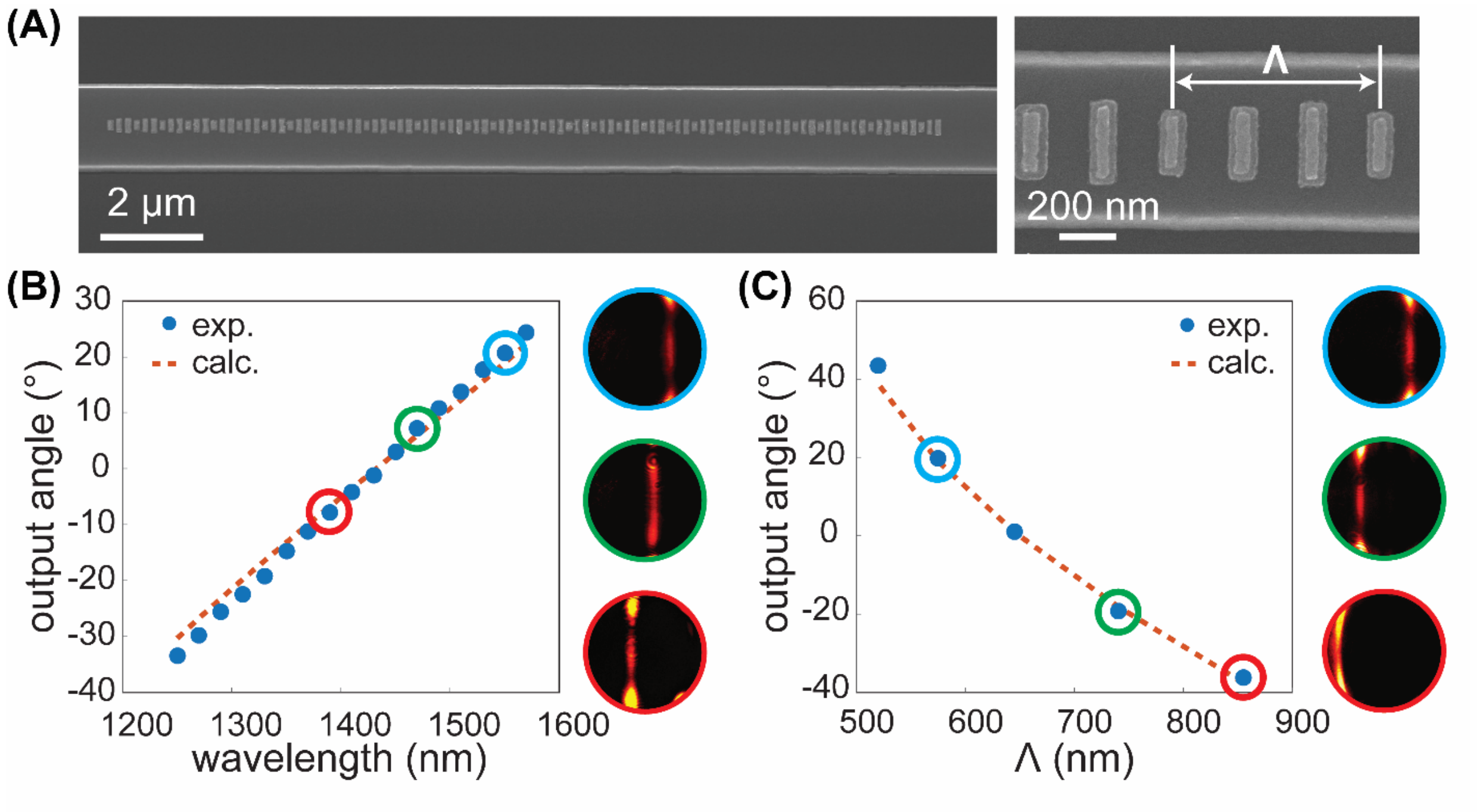

**Fig. 3. Demonstration of off-chip beam steering with guided-wave-driven metasurfaces.** (**A**) Field emission scanning electron microscope (FESEM) images of a guided-wave-driven metasurface on a silicon waveguide (220 nm thick and 600 nm wide). Each supercell consists of three meta-atoms as depicted in Fig. 2B. (**B**) Output beam angle versus the incident guided wave wavelength with supercell size Λ = 575 nm measured by our Fourier-space imaging system (Fig. S1). The blue dots and the red dashed line depicts the experimentally measured and the simulated data, respectively. Three typical Fourier-space images of the extracted free-space light corresponding to the circled data points are shown on the right. The horizontal and vertical axes represent $k_x$ and $k_y$ respectively. An objective with numerical aperture (NA) of 0.95 was used in the measurements. (**C**) Output beam angle versus the supercell size at 1550 nm wavelength. The blue dots and the red dashed line depicts the experimentally measured and the simulated data, respectively. Similar to (B), three typical Fourier-space images are shown on the right.

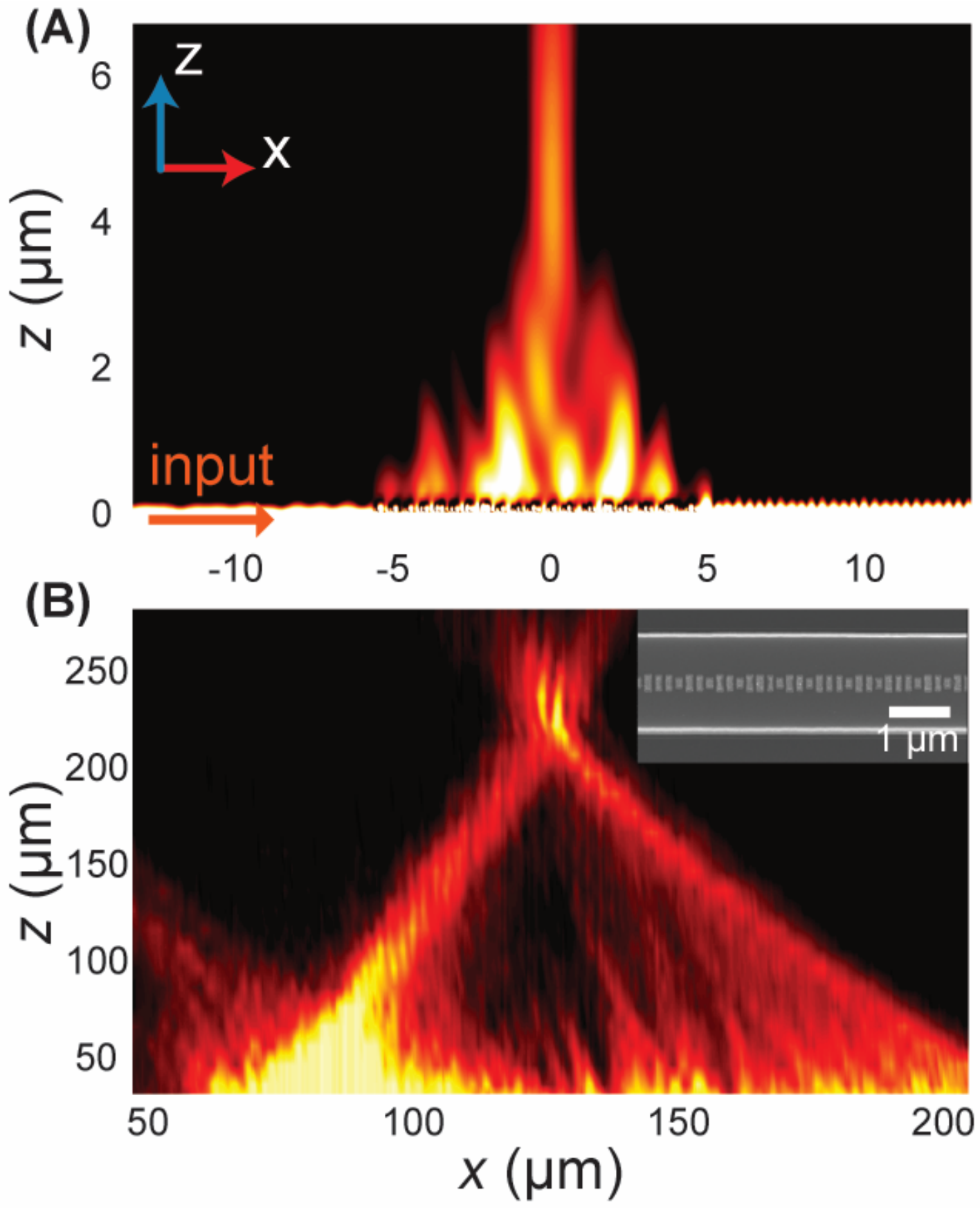

**Fig. 4. Demonstration of off-chip light focusing with a guided-wave-driven metalens.** (**A**) Simulated electric field distribution above a guided-wave-driven metalens on a silicon waveguide (500 nm thick and 1.5 μm wide). The extracted light converged at the designed focal point (5 μm above the waveguide) at 1550 nm wavelength. (**B**) Experimentally measured intensity profile of the focusing effect of a fabricated device. The inset shows an FESEM image of the metasurface region. The designed focal length is 225 μm.

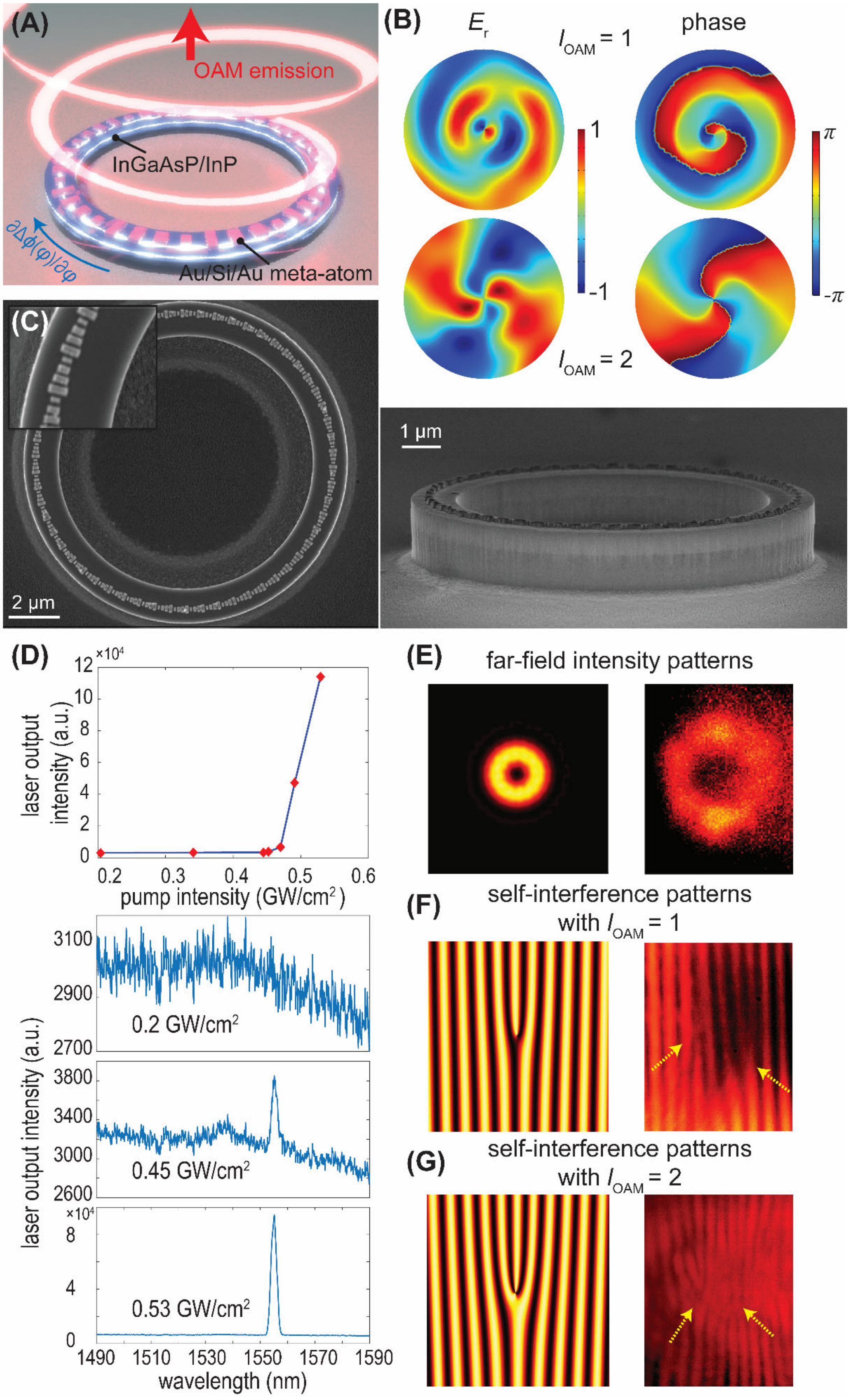

**Fig. 5. A photonic integrated OAM laser based on the guided-wave-driven metasurface incorporated on an InGaAsP/InP micoring resonator.** (**A**) A schematic of a micro-ring OAM laser enabled by the guided-wave-driven metasurface. Unidirectional phase modulation provided by the metasurface breaks the degeneracy of the CCW and CW WGMs inside the micro-ring resonator, leading to a selective OAM radiation. (**B**) Simulated electric field (radial component) and phase distribution of emitted wave with different numbers of metasurface supercells *N*. The azimuthal order of the WGM is $M = 59$ at the resonant wavelength of 1550 nm, and the number of supercells is $N = 58$ (top row) and $N = 57$ (bottom row). The resulting topological charge of the OAM radiation can be seen by the number of $2\pi$ phase evolution along the circumference, which is +1 (top row) and +2 (bottom row), respectively. (**C**) FESEM images of a fabricated device. The diameter of the micro-ring is 9 μm and the width is 1.1 μm, and it consists of a 500-nm InGaAsP MQW layer and a 1-μm InP layer. A supercell of the metasurface consists of four Au/Si/Au meta-atoms, which provides the extracted wave with abrupt phase shifts from 0 to $2\pi$. The total number of supercell on the micro-ring is $N = 58$. (**D**) Light-light curve of the micro-ring laser (top row), which shows a lasing threshold of about 0.47 GW/cm$^2$ at 1555 nm wavelength. Three emission spectrum corresponding to different stages – photoluminescence, amplified spontaneous emission, and lasing – of the laser are shown from 2$^{nd}$ to the last row. (**E**) Far-field intensity distribution of the OAM laser radiation captured by an infrared camera (right panel), which matches well with the simulated one (left panel). Both figures show an annular shape. (**F**) and (**G**) Calculated (left panels) and measured (right panels) self-interference patterns of OAM laser radiation. The double-fork (F) and triple-fork (G) in the fringe patterns confirmed that the resulting OAM emission has a topological charge of +1 (F) and +2 (G), respectively.

**Data Availability**

The data that support the findings of this study are available from the corresponding author upon reasonable request.

**Code Availability**

The MATLAB codes used in this study are available from the corresponding author upon reasonable request.

Supplementary Material for

# Molding Free-Space Light with Guided-Wave-Driven Metasurfaces

**Authors:** Xuexue Guo, Yimin Ding, Xi Chen, Yao Duan, and Xingjie Ni*

**Affiliations:** Department of Electrical Engineering, Pennsylvania State University, University Park, PA 16802, USA.

**Correspondence to:** xingjie@psu.edu

## 1. Device fabrication

The samples were fabricated on a commercially available silicon-on-insulator wafer with 220-nm-thick (for beam steering experiments) and 500-nm-thick (for light focusing experiments) Si device layer and 3-µm buried silicon dioxide. The wafer was cleaned by sonication in acetone and IPA for 3 minutes, respectively. Alignment marker was defined by electron beam lithography with 100 kV beam (Vistec EBPG5200) followed by evaporation of 50 nm Au with a 5-nm-thick Ti adhesion layer (Kurt J. Lesker Lab-18) and lift-off. Then negative resist Fox-16 (Dow Corning Corp.) was spin-coated and prebaked at 100 °C for 4 minutes. The waveguide pattern was written followed by development in CD-26 developer (MicroChem) for 25 minutes to reduce proximity effect. Chlorine-based inductively coupled reactive ion etching (ICP-RIE) was used to etch crystalline Si with FOX-16 resist as mask (Plasma-Therm Versalock 700). The sample was immersed in buffered oxide etchant for 20 seconds followed by water rinse to remove the remaining mask. ZEP 520A (Zeon) was spin-coated on the sample and soft-baked at 180 °C for 3 minutes. A second-step electron beam lithography was conducted to define the metasurface layer on top of the waveguide with precise alignment. The exposed sample was developed in N-amyl-acetate for 3 minutes followed by MIBK:IPA immersion for 1 minute. Au/SiO2/Au films were subsequently deposited using an electron beam evaporation system (Semicore). The pattern was then lifted off in 1165 remover (MicroChem) at 85 °C in water bath for 2 hours. The sample was finally diced along the input port of waveguide for measurement.

The OAM micro-ring laser was fabricated on InGaAsP (500 nm, multi-quantum-well layer) / InP substrate. First the micro-ring resonator was defined by electron beam lithography with FOX-16 negative resist. The resist acted as an etch mask in the $BCl_3$ based ICP-RIE process. Then the sample was immersed in buffered oxide etchant to remove the mask. A second-step electron beam lithography using ZEP 520A resist was performed with precise alignment to define the metasurface layer on top of the micro-ring resonator. A sequential electron beam evaporation was done to deposit Au/Si/Au films, followed by a standard lift-off process in 1165 remover at 85 °C in water bath for 2 hours.

## 2. Simulation methods

Numerical simulations were carried out using a commercially available finite element method (FEM) solver package – COMSOL Multiphysics. Third-order finite elements and at least 10 mesh steps per wavelength were used to ensure the accuracy of the calculated results. We simulated individual Au/$SiO_2$/Au meta-atoms first. We used an eigen-mode solver to find the $TE_{00}$ mode of the silicon waveguide as well as its modal index at 1550 nm wavelength. Then this modal index was used in the model to further calculate the phase and amplitude of the extracted light by monitoring the field at a few wavelengths over the waveguide. We swept the geometrical parameters of the meta-atoms to get the phase and amplitude maps/contours (Fig. 2B). The trapezoidal shape of meta-atoms resulted from our nanofabrication was also taken into account in our model to get accurate design parameters.

In order to simulate the beam steering, a full device model that consists of an array of meta-atoms placed on top of Si waveguide was established. The meta-atoms were distributed along the waveguide so that they formed a linear phase gradient (Fig. 2D). In a similar fashion, an array of meta-atoms fulfilling the spatial phase distribution of a lens were placed on top of Si waveguide to simulate the light focusing effect (Fig. 4A).

We used a similar method to calculate the phase and amplitude of the extracted light from Au/Si/Au meta-atoms on top of InGaAsP/InP waveguide (Fig. S3A). A device-level model of metasurface incorporated micro-ring resonator was constructed to simulate the OAM radiation. Four Au/Si/Au meta-atoms selected from calculated phase and amplitude maps/contours (Fig. S3A) were used to construct a supercell. 58 supercells ($N = 58$) were placed on top of the resonator. Using the WGM with $M = 59$, we achieved OAM radiation with topological charge of +1 according to Eq. (4). Fig. S3B illustrates the electric field distribution of $TE_{00}$ mode which shows the typical standing wave pattern formed by the two counter-propagating (CW and CCW) WGMs. The extracted light in free space carries OAM as shown in Fig. 5B.

## 3. Characterization of beam steering and focusing on waveguide-fed metasurfaces

We characterized our waveguide-fed metasurfaces with beam steering and focusing functions using the optical setup shown in Fig. S1. A free-space laser beam output from a Ti:Sapphire laser pumped optical parametric oscillator (OPO) was coupled into a commercially available tapered lensed single-mode fiber. The focused laser beam from the tapered fiber was coupled into the input port of our fabricated ridge waveguide sample in end-fire manner by using a three dimensional (3D) translational stage. The coupled-in light propagated through a triangle taper linking the input port and the single-mode waveguide, during which the high-order modes vanished and only fundamental transverse electrical mode survived. The light scattered into free space by metasurfaces on top of the single-mode waveguide was collected by an objective (N.A. = 0.95) and then transmitted through a tube lens. Part of the light was reflected by a beamsplitter for real-space imaging. And the light transmitted through the beamsplitter was focused by a Bertrand lens to form *Fourier-space* image. In off-chip beam-steering measurement, the laser wavelength was tuned using the OPO to acquire wavelength-dependent beam steering angles. In addition, Fourier-space images were taken by coupling 1550nm laser beam into samples with different supercell periods. After that, the scattering angles were extracted from the Fourier-space images calibrated by a ruled reflective grating (grooves density of 600/mm). In off-chip focusing measurement, the samples were mounted on a 3D translational stage with a high-resolution piezo-controlled actuator in *z*-direction. By moving the *z* stage, the real-space images were taken at different distance from the waveguide plane and a 3D intensity distribution was reconstructed.

## 4. Characterization of OAM lasing on waveguide-fed metasurfaces integrated InGaAsP/InP micro-ring resonator

Fig. S4 shows the experimental setup to observe the lasing spectra and to confirm the OAM properties of the laser radiation. In our experiment, a femtosecond pulsed laser (~140 fs, repetition rate 80 MHz) at 900 nm wavelength was reflected by a dichroic mirror and then focused by a Newport 20X objective (NA = 0.40) onto the micro-ring resonator. The pump power was controlled by a circular variable neutral density filter and monitored by a power meter. The lasing emission was collected by the same objective and then transmitted through the dichroic mirror and detected by a spectrometer (Horiba), a far-field imaging system and a Michelson interferometry setup. With a flip mirror to switch the paths, the laser emission was either sent into the spectrometer/imaging system or the interferometry setup. In the interferometry setup, the laser emission was split into two beams by a pellicle beam splitter, and then recombined with an off-center beam overlap to form an interference pattern recorded by an infrared camera. A delay line was used in order to balance the optical path lengths of the two arms.

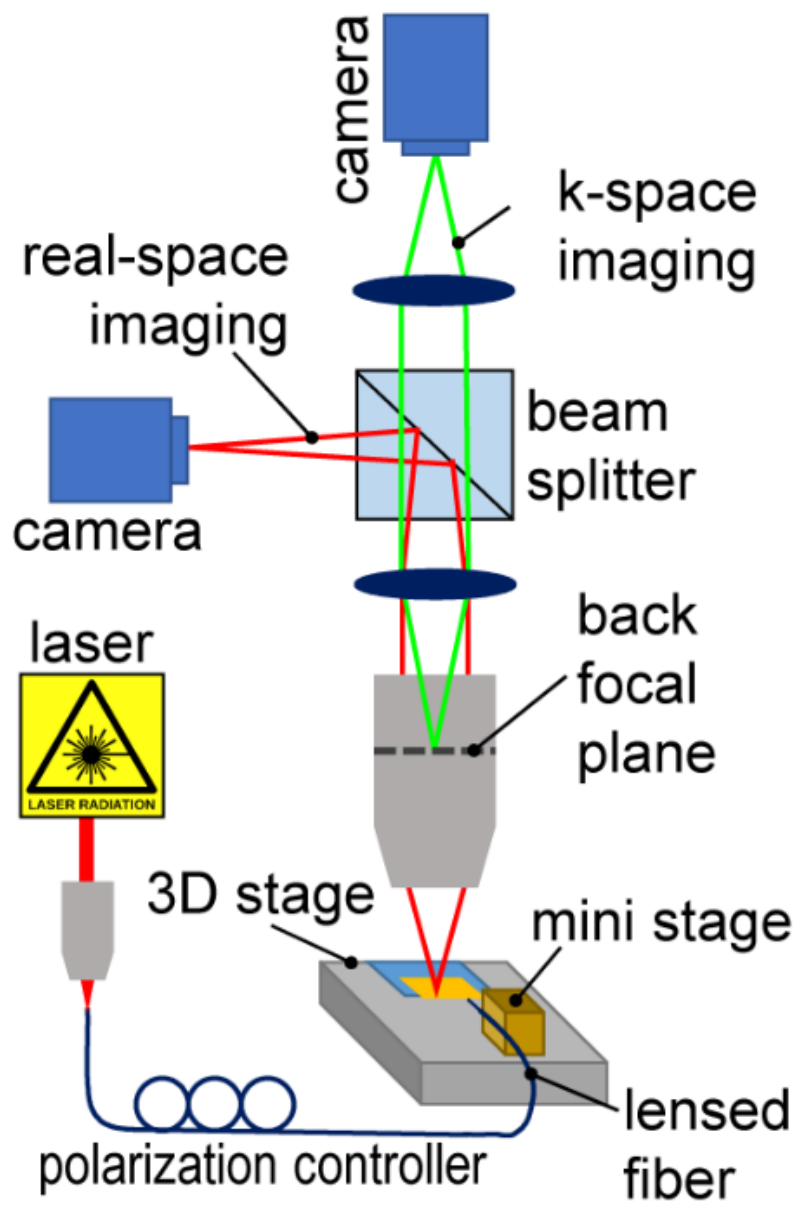

**Fig. S1**. A schematic of the experimental setup for the off-chip beam steering and focusing measurements. A free-space laser beam emitted from a Ti:Sapphire laser pumped OPO was coupled into a tapered lensed single-mode fiber and then to the input port of the sample. The extracted light in free space was collected by an objective (NA = 0.95) and then transmitted through a tube lens. The light was partially reflected by a non-polarizing beam splitter for real-space imaging. And the rest light was transmitted through the beam splitter and was focused by a Bertrand lens to form *Fourier-space* images.

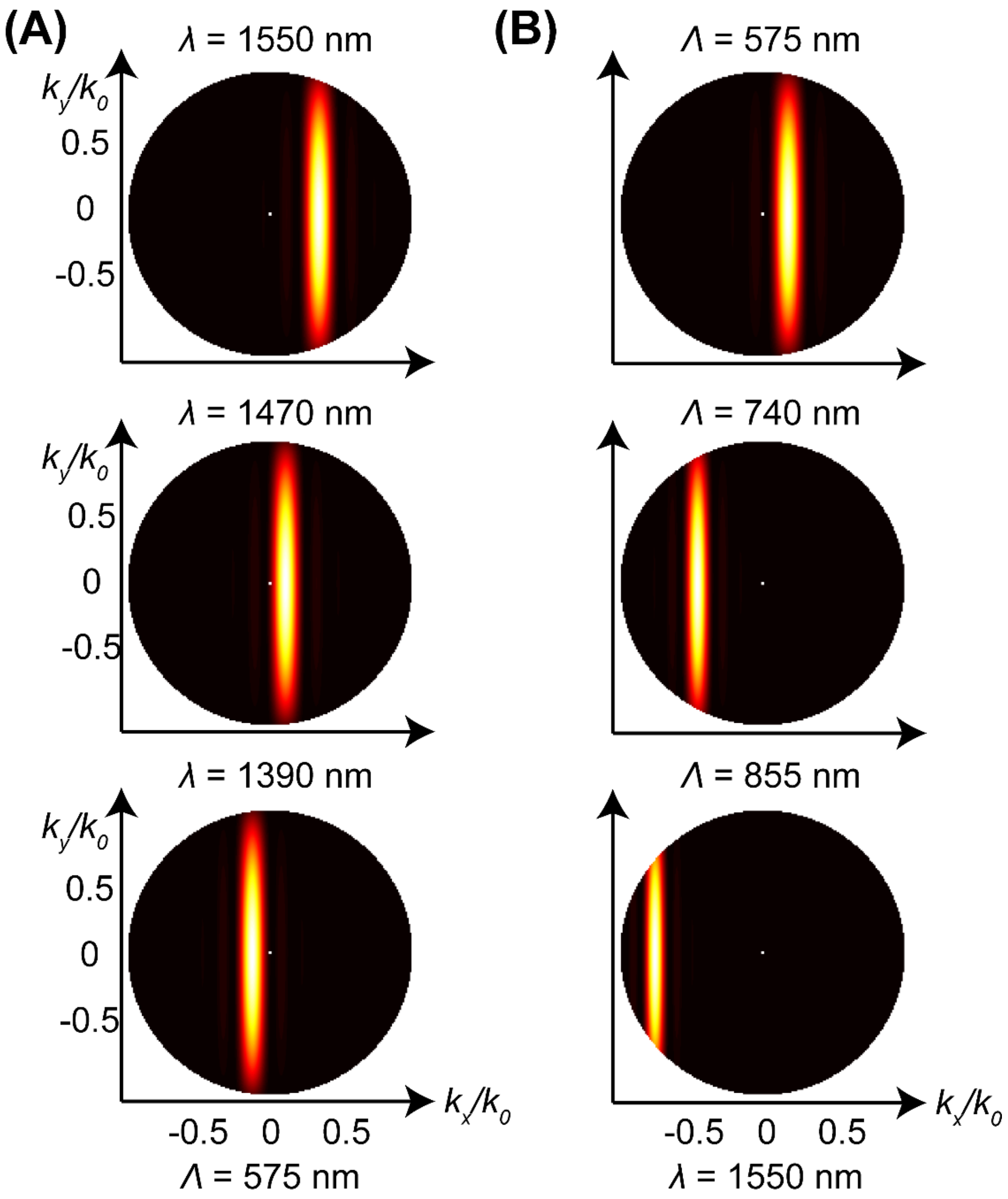

**Fig. S2** (**A**) Simulated Fourier-space images of the extracted light at different wavelengths from a metasurface with supercell size of 575 nm. (**B**) Simulated Fourier-space images of the extracted light from metasurfaces with different supercell sizes at 1550 nm wavelength.

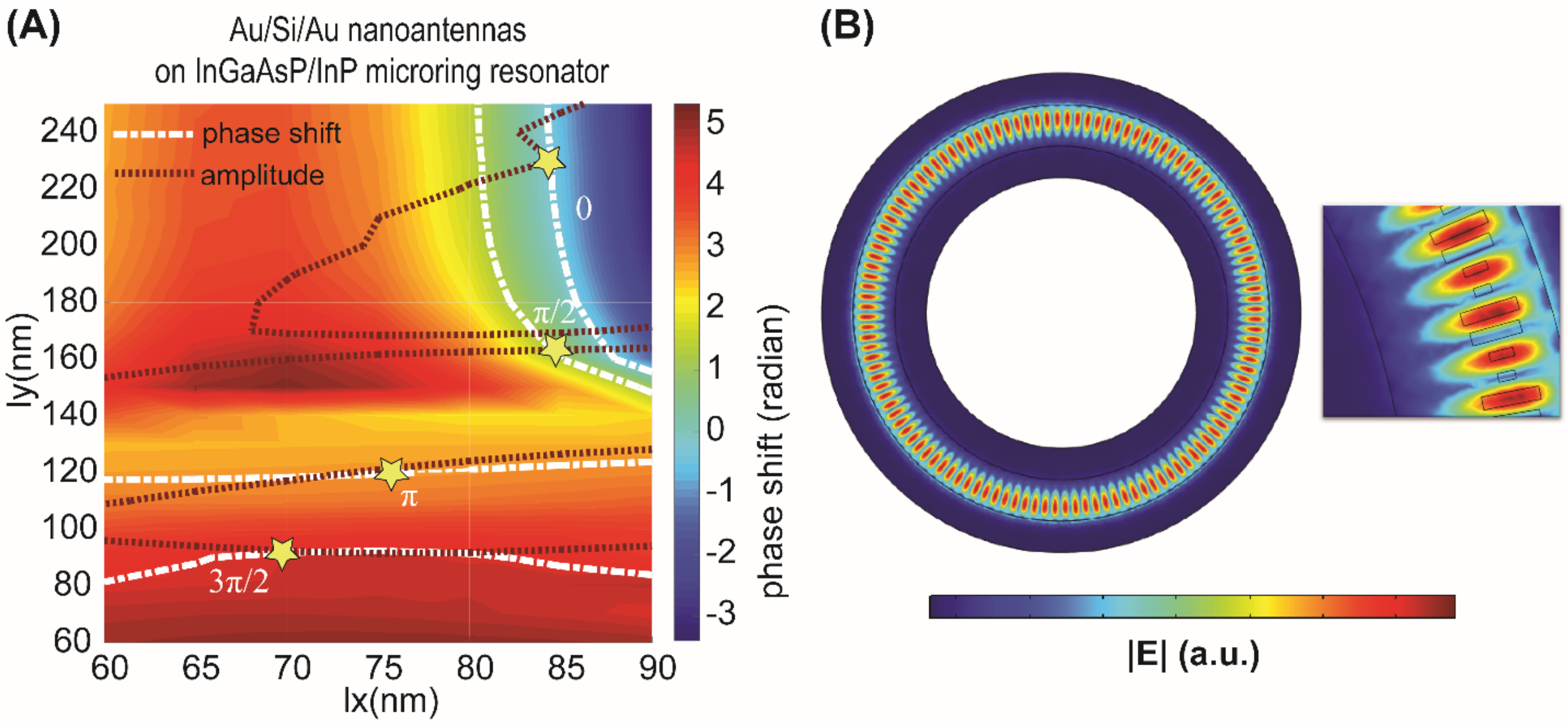

**Fig. S3** (**A**) A pseudo-color map of the simulated abrupt phase shifts overlaid with amplitude contours generated by the Au/Si/Au meta-atoms with different lengths ($l_x$) and widths ($l_y$). Four meta-atom designs (marked by the yellow stars) with a constant phase shift difference of π/2 were selected to construct the metasurface supercell. The black dashed line is the contour of the extracted electric field amplitude of $2 \times 10^5$ V/m. (**B**) The simulated electric field distribution of the micro-ring resonator (diameter = 9 μm, width = 1.1 μm and height = 1.5 μm) with WGM order $M$ = 59. A close-up view of one segment of the micro-ring shows a good spatial overlap between the waveguide mode and the meta-atoms.

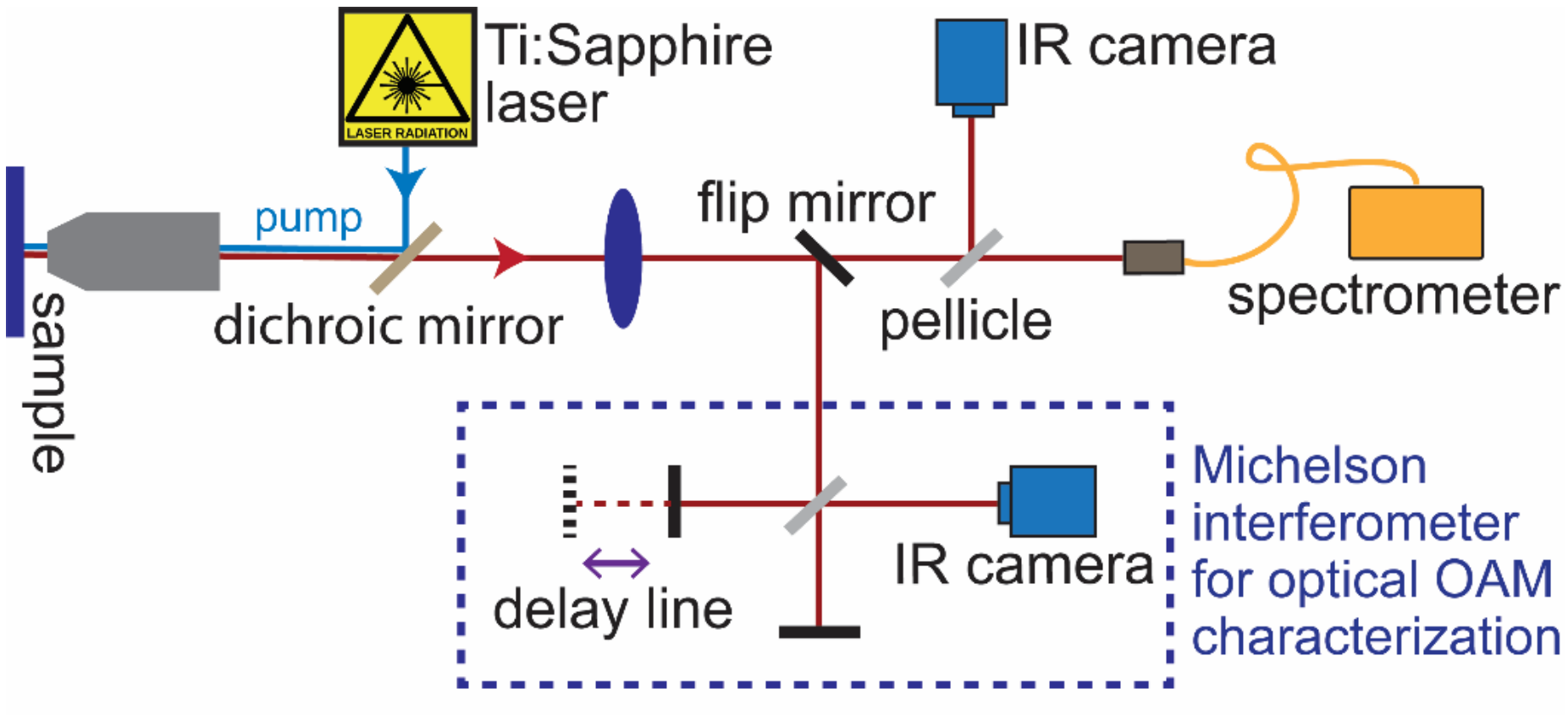

**Fig. S4.** A schematic of the experimental setup for characterizing the OAM laser emission. A femtosecond pulsed pump laser (~140 fs, repetition rate 80 MHz) at 900 nm wavelength was reflected by a dichroic mirror and then focused by a Newport 20X objective (NA = 0.40) onto the micro-ring resonator. The lasing emission was collected by the same objective and then transmitted through the dichroic mirror to be detected by a spectrometer, a far-field imaging system and a Michelson interferometry setup. A flip mirror was used to switch the paths.